\documentclass[twoside,twocolumn,9pt]{article}
\usepackage{extsizes}
\usepackage[super,sort&compress,comma]{natbib}
\usepackage[version=3]{mhchem}
\usepackage[left=1.5cm, right=1.5cm, top=1.785cm, bottom=2.0cm]{geometry}
\usepackage{balance}
\usepackage{mathptmx}
\usepackage{sectsty}
\usepackage{graphicx}
\usepackage{lastpage}
\usepackage[format=plain,justification=justified,singlelinecheck=false,font={stretch=1.125,small,sf},labelfont=bf,labelsep=space]{caption}
\usepackage{float}
\usepackage{fancyhdr}
\usepackage{fnpos}
\usepackage[english]{babel}
\addto{\captionsenglish}{%
  
}
\usepackage{array}
\usepackage{droidsans}
\usepackage{charter}
\usepackage[T1]{fontenc}
\usepackage[usenames,dvipsnames]{xcolor}
\usepackage{setspace}
\usepackage[compact]{titlesec}
\usepackage{hyperref}
\usepackage{amsmath}
\usepackage{siunitx}

\definecolor{cream}{RGB}{222,217,201}

\begin{document}

\pagestyle{fancy}
\thispagestyle{plain}
\fancypagestyle{plain}{
\renewcommand{\headrulewidth}{0pt}
}

\makeFNbottom
\makeatletter
\renewcommand\LARGE{\@setfontsize\LARGE{15pt}{17}}
\renewcommand\Large{\@setfontsize\Large{12pt}{14}}
\renewcommand\large{\@setfontsize\large{10pt}{12}}
\renewcommand\footnotesize{\@setfontsize\footnotesize{7pt}{10}}
\makeatother

\renewcommand{\thefootnote}{\fnsymbol{footnote}}
\renewcommand\footnoterule{\vspace*{1pt}%
\color{cream}\hrule width 3.5in height 0.4pt \color{black}\vspace*{5pt}}
\setcounter{secnumdepth}{5}

\makeatletter
\renewcommand\@biblabel[1]{#1}
\renewcommand\@makefntext[1]%
{\noindent\makebox[0pt][r]{\@thefnmark\,}#1}
\makeatother
\renewcommand{\figurename}{\small{Fig.}~}
\sectionfont{\sffamily\Large}
\subsectionfont{\normalsize}
\subsubsectionfont{\bf}
\setstretch{1.125} 
\setlength{\skip\footins}{0.8cm}
\setlength{\footnotesep}{0.25cm}
\setlength{\jot}{10pt}
\titlespacing*{\section}{0pt}{4pt}{4pt}
\titlespacing*{\subsection}{0pt}{15pt}{1pt}

\fancyfoot{}
\fancyfoot[LO,RE]{\vspace{-7.1pt}\includegraphics[height=9pt]{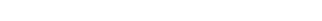}}
\fancyfoot[CO]{\vspace{-7.1pt}\hspace{13.2cm}\includegraphics{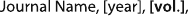}}
\fancyfoot[CE]{\vspace{-7.2pt}\hspace{-14.2cm}\includegraphics{head_foot/RF}}
\fancyfoot[RO]{\footnotesize{\sffamily{1--\pageref{LastPage} ~\textbar  \hspace{2pt}\thepage}}}
\fancyfoot[LE]{\footnotesize{\sffamily{\thepage~\textbar\hspace{3.45cm} 1--\pageref{LastPage}}}}
\fancyhead{}
\renewcommand{\headrulewidth}{0pt}
\renewcommand{\footrulewidth}{0pt}
\setlength{\arrayrulewidth}{1pt}
\setlength{\columnsep}{6.5mm}
\setlength\bibsep{1pt}

\makeatletter
\newlength{\figrulesep}
\setlength{\figrulesep}{0.5\textfloatsep}

\newcommand{\topfigrule}{\vspace*{-1pt}%
\noindent{\color{cream}\rule[-\figrulesep]{\columnwidth}{1.5pt}} }

\newcommand{\botfigrule}{\vspace*{-2pt}%
\noindent{\color{cream}\rule[\figrulesep]{\columnwidth}{1.5pt}} }

\newcommand{\dblfigrule}{\vspace*{-1pt}%
\noindent{\color{cream}\rule[-\figrulesep]{\textwidth}{1.5pt}} }

\makeatother

\twocolumn[
  \begin{@twocolumnfalse}
{\includegraphics[height=30pt]{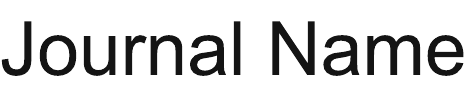}\hfill\raisebox{0pt}[0pt][0pt]{\includegraphics[height=55pt]{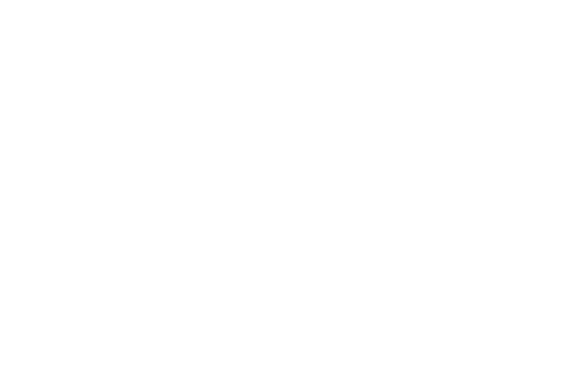}}\\[1ex]
\includegraphics[width=18.5cm]{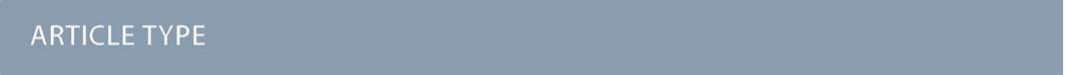}}\par
\vspace{1em}
\sffamily
\begin{tabular}{m{4.5cm} p{13.5cm} }

\includegraphics{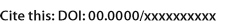} & \noindent\LARGE{\textbf{Time-resolved solvation dynamics of Li$^+$, Na$^+$ and K$^+$ ions in liquid helium nanodroplets}} \\
\vspace{0.3cm} & \vspace{0.3cm} \\

 & \noindent\large{Jeppe K. Christensen,\textit{$^{a}$} Simon H. Albrechtsen,\textit{$^{a}$} Christian E. Petersen,\textit{$^{b}$} Constant A. Schouder,\textit{$^{c}$}
Iker S\'anchez-P\'erez,\textit{$^{d}$} Pedro Javier Carchi-Villalta,\textit{$^{d}$} Massimiliano Bartolomei,\textit{$^{d}$} Fernando Pirani,\textit{$^{e}$} Tomás González-Lezana,\textit{$^{d}$} and Henrik Stapelfeldt\textit{$^{a}$}$^{\ast}$} \\

\includegraphics{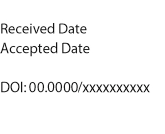} & \noindent\normalsize{In 2023, ultrafast pump-probe spectrocopy was used to record the solvation dynamics of a single Na$^+$ ion in a liquid helium droplet, atom-by-atom and with femtosecond time resolution [Albrechtsen \textit{et al., Nature}, 2023, \textbf{623}, 319]. Subsequently, theoretical studies showed that other alkali ions solvate in a similar manner but no experimental results were reported so far. Here, we extend the previous measurement on Na$^+$ to Li$^+$ and K$^+$ ions. A pump pulse selectively ionizes an alkali atom, initially residing at the droplet surface, and the ensuing solvation dynamics of the formed alkali cation, \ce{Ak+}, is followed by ionizing a Xe atom, located in the droplet interior, and recording the yields of Ak$^+$He$_n$ ions expelled from the droplet as a function of the pump-probe pulse delay. We find that Li$^+$, Na$^+$ and K$^+$ ions solvate with a binding rate of 1.8~$\pm$~0.1, 1.8~$\pm$~0.1 and 1.7~$\pm$~0.1 He per ps, respectively. Furthermore, by comparing the number distribution of the Ak$^+$He$_n$ ion yields to the evaporation energies of these ion--He complexes, obtained by Path Integral Monte Carlo calculations, we identify signatures of the first solvation shells of \ce{Li+}, \ce{Na+} and \ce{K+}. Lastly, we determine the time-dependent dissipation of the solvation energy from the vicinity of the three alkali ion species and find that the rate is highest (lowest) for \ce{Li+} (\ce{K+}).
} \\

\end{tabular}

 \end{@twocolumnfalse} \vspace{0.6cm}

  ]

\renewcommand*\rmdefault{bch}\normalfont\upshape
\rmfamily
\section*{}
\vspace{-1cm}


\footnotetext{\textit{$^{a}$Department of Chemistry, Aarhus University, Langelandsgade 140, DK-8000 Aarhus C, Denmark. E-mail: henriks@chem.au.dk}}
\footnotetext{\textit{$^{b}$Department of Physics and Astronomy, Aarhus University, Ny Munkegade 120, DK-8000 Aarhus C, Denmark.}}
\footnotetext{\textit{$^{c}$Université Paris-Saclay, CNRS, Institut des Sciences Moléculaires d'Orsay, 91405 Orsay, France.}}
\footnotetext{\textit{$^{d}$Instituto de Física Fundamental, CSIC (IFF-CSIC), Serrano 123, 28006 Madrid, Spain.}}
\footnotetext{\textit{$^{e}$Dipartimento di Chimica, Biologia e Biotecnologie, Universit\'a di Perugia, 06123 Perugia, Italy.}}


\section*{Introduction}
Ions and their solvation structures play a crucial role in both chemical and biological systems,\cite{marcus_ions_2015} where an important elemental group is the alkali metal ions.\cite{astrid_alkali_2016} The solvation structure of Li$^+$ ions in Li-ion batteries is important for the battery efficiency,\cite{wang_structural_2022} the hydration number of the Na$^+$ ion significantly changes its transport properties\cite{peng_effect_2018} and the solvation and desolvation dynamics of the K$^+$ ion are central to the functioning of cellular potassium channels.\cite{morais-cabral_energetic_2001} It is therefore of significant interest to understand the fundamentals of the cation solvation process and how this depends on the different alkali species. In practice, this requires experiments capable of following the transfer of an ion from an initial unsolvated state to a (fully) solvated state in a liquid i.e. the process where solvent molecules -- or atoms -- are attracted to and bind to the ion. This is an experimentally challenging task due to the need for methods sensitive to the instantaneous solvation state,\cite{yu_solvation_2025} namely how many solvent particles the ion has bound at a particular time. Furthermore, the experiment should operate with picosecond or even sub-picosecond time resolution to capture the binding dynamics of individual solvent particles.

A model system useful for addressing these experimental challenges is superfluid liquid helium droplets. Liquid helium can readily dissolve cations including the alkali metal ions\cite{glaberson_impurity_1975,atkins_ions_1959,tabbert_optical_1997} and nanoscopic droplets of liquid helium have been used extensively to host single atoms, molecules, clusters or complexes in a cold, weakly perturbing environment.\cite{toennies_superfluid_2004,choi_infrared_2006,mauracher_cold_2018,stadlhofer_real-time_2025} In particular, neutral alkali metal atoms are known to be weakly bound on the surface of the droplets,\cite{stienkemeier_spectroscopy_1996,barranco_helium_2006} from which they solvate and move into the droplet upon ionization\cite{theisen_forming_2010,zhang_communication:_2012,garcia-alfonso_time-resolved_2024} -- an example of ion solvation in an atomic bath.\cite{chowdhury_ion_2024} During the solvation, the ion breaks the local superfluidity of the droplet by forming a solvation shell,\cite{gonzalez-lezana_solvation_2020, matos_transitional_2025,yanes_microsolvation_2025} where the charge density and ionic size of the alkali metal ion plays an important role for the size and structure of the shells. Thus, ionization of an alkali atom at the droplet surface with a femtosecond laser pulse opens unique opportunities for real-time exploration of the solvation process of the corresponding alkali ion as it starts essentially from a gas phase, i.e. unsolvated state. Since helium is transparent well into the deep UV, the laser pulse does not perturb the surrounding environment.

Alkali metal ions solvated in liquid helium droplets have been extensively studied in static and dynamic measurements. For instance, electron impact ionization\cite{an_der_lan_solvation_2012} and femtosecond photoionization\cite{muller_alkali-helium_2009} of surface-located alkali metal atoms were both used to measure mass spectra of Ak$^+$He$_N$ clusters, revealing particularly stable structures that are interpreted as closed solvation shells. Theoretical studies\cite{gonzalez-lezana_solvation_2020,rastogi_lithium_2018} have also shown how the binding energy of individual helium atoms changes as a function of the number of helium atoms attached to the Ak$^+$He$_N$ ion. Here, steep decreases in the binding energy have been observed at specific values of $N$ indicating the closure of solvation shells. Concerning the solvation dynamics of alkali ions, several results were reported using different methods.\cite{leal_picosecond_2014,leal_dynamics_2016,garcia-alfonso_time-resolved_2024,garcia-alfonso_time-resolved_2025,calvo_concurrent_2024,calvo_time-dependent_2025} Time-dependent liquid $^4$He density-functional theory predicts that all alkali metal species, Li through Cs, solvate into the helium droplet upon ionisation.\cite{garcia-alfonso_time-resolved_2024,garcia-alfonso_time-resolved_2025} For \ce{Li+}, a binding rate of 1.46 He per ps has been found,\cite{garcia-alfonso_time-resolved_2024} while \ce{Na+} and \ce{K+} were found to bind with a rate of 1.3 and 1.0 He per ps.\cite{garcia-alfonso_time-resolved_2025} These studies also indicate that the solvation process can be described as Poissonian for the first several helium atoms. Another approach,\cite{calvo_concurrent_2024} utilising ring-polymer molecular dynamics, has similarly been employed to show how a Na$^+$ solvates into the droplet upon ionisation. These simulations are atomistic, so the coordination number of He around Na$^+$ can be followed as a function of time, where a stable structure of N = 10 was found, interpreted as the first solvation shell.

Recently, we showed experimentally that the initial solvation dynamics of a Na$^+$ ion sinking into a helium droplet can be monitored on the femtosecond timescale by using a pump-probe laser scheme.\cite{albrechtsen_observing_2023, albrechtsen_femtosecond-and-atom-resolved_2025} In that work, a helium droplet was doubly doped with a Na atom at the surface and a Xe atom at the center of the droplet. The pump laser pulse selectively ionised the Na atom, thereby initiating solvation of Na$^+$. At a well-defined time delay, the probe pulse ionised Xe, which repels Na$^+$ and any He atoms attached. Detection of these Na$^+$He$_n$ ions, as a function of the time delay between the two laser pulses, allowed us to determine the solvation dynamics atom-by-atom. We found that the solvation process for the first five He atoms can be described as Poissonian with a binding rate that depends on the droplet size and is $\sim$~1.8 atoms/ps for a droplet containing 5200 atoms. Using evaporation energies obtained by Path Integral Monte Carlo (PIMC) simulations on a specifically calculated new potential energy surface (PES) describing the interactions between the atoms in the Na$^+$He$_n$ ions, the time-dependent dissipation of energy from the region around the sodium ion was determined. Here, we extend the studies to two other alkali metals, Li and K. The purpose of this work is twofold: first, to establish the pump-probe technique developed for Na$^+$ as a more general method for monitoring solvation dynamics of surface-bound species on helium droplets. Secondly, to investigate how the solvation dynamics of Li$^+$ and K$^+$ ions differ from that of Na$^+$ and how these compare to theoretical results.

\section*{Methods}
\subsection*{Experiment}
The experimental setup has previously been explained in detail\cite{albrechtsen_femtosecond-and-atom-resolved_2025} and will be briefly stated here with changes between the alkali species noted.
A beam of liquid helium droplets is formed by expanding 50 bar of 99.9999\% purity He gas through a 5 µm round orifice into vacuum. The expansion nozzle is cooled to 18 K in all measurements, resulting in a distribution of helium droplet sizes with an average number of helium atoms $\langle N_D\rangle$ = 5200. The helium droplets have an expected temperature of 0.37 K.\cite{hartmann_rotationally_1995} The helium droplet beam enters a second vacuum chamber through a 2 mm-diameter skimmer, where the droplets are doubly doped with a single xenon atom and a single alkali atom (Li, Na, K). The gaseous Xe sample is introduced through a leak valve, while the alkali metal vapor is formed by heating a sample of the metal. The alkali metal samples were heated to 400 °C, 180 °C and 75 °C for Li, Na, and K, respectively. The vapor pressures of both elements were optimized for single doping of the helium droplets, but the formation of some larger clusters is unavoidable due to the statistical nature of the doping process.\cite{toennies_superfluid_2004}
The doped droplet beam passes through another 2 mm-diameter skimmer into the target vacuum chamber, where it is crossed by two focused laser beams in the center of a velocity map imaging (VMI) spectrometer.\cite{chandler_twodimensional_1987,eppink_velocity_1997}

The results on Li and Na were obtained using laser pulses from a Spectra-Physics Spitfire Ace (800 nm, 50 fs, 4 mJ, 1 kHz) laser system and results on K were obtained using a Coherent Astrella HE (800 nm, 35 fs, 2 mJ, 5 kHz) laser system with specific parameters shown in Table \ref{tbl:laserparams_pump}.
\begin{table}[h]
\small
  \caption{\ Overview of the laser parameters of the pump and probe laser pulses used in the experiments on Li, Na and K, including central wavelength of the laser pulse ($\lambda$), pulse energy ($E_{\text{pulse}}$), pulse duration ($\tau$), focused spot size ($w_0$) and peak intensity (I$_0$)}\label{tbl:laserparams_pump}
  \begin{tabular*}{\columnwidth}{@{\extracolsep{\fill}}llllll}
    Pump pulse\\
    \hline
    Alkali & $\lambda$ (nm) & $E_{\text{pulse}}$ (µJ) & $\tau$ (fs) & w$_0$ (µm) & I$_0$ (W/cm$^2$) \\
    \hline
    Li & 400 & 25 & 100 & 24 & 2.6$\times$10$^{13}$ \\
    Na & 800 & 35 & 65 & 24 & 5.8$\times$10$^{13}$ \\
    K & 800 & 6 & 41 & 44 & 4.5$\times$10$^{12}$ \\
    \hline
    \\
    Probe pulse\\
    \hline
    Alkali & $\lambda$ (nm) & $E_{\text{pulse}}$ (µJ) & $\tau$ (fs) & w$_0$ (µm) & I$_0$ (W/cm$^2$) \\
    \hline
    Li & 800 & 25 & 63 & 21 & 1.2$\times$10$^{14}$ \\
    Na & 400 & 35 & 100 & 16 & 9.1$\times$10$^{13}$ \\
    K & 400 & 26 & 51 & 16 & 1.0$\times$10$^{14}$ \\
    \hline

  \end{tabular*}
\end{table}
The color and intensity of the pump pulse were chosen to efficiently ionize the alkali metal atom while avoiding ionization of Xe. The probe pulse intensity was chosen to efficiently ionize Xe to Xe$^+$ while avoiding double ionization of Xe and saturation of the detector from Xe$^+$He$_k$ ions.
Both laser pulses were linearly polarized in the plane of the position-sensitive detector at the end of the VMI spectrometer. The detector consists of a pair of microchannel plates backed by a P47 phosphor screen, which is imaged by a TPX3Cam.\cite{fisher-levine_timepixcam_2016,zhao_coincidence_2017,nomerotski_imaging_2019} The TPX3Cam records the position and time-of-arrival for each lit up pixel independently and these values are converted to a time-of-flight (ToF) and a pair of coordinates, $(x, y)$, representing the pixel. The conversion includes correcting for time-walk effects,\cite{bromberger_shot-by-shot_2022}\cite{pitters_time_2019} clustering and centroiding the data using the DBSCAN algorithm. The ToF values are calibrated using known molecular species to mass-over-charge, $m/q$, values and the $(x, y)$ coordinates are calibrated to the projected $(v_x, v_y)$ velocities using the known velocities of the Ak$^+$ ions originating from the laser-induced Coulomb explosion of the relevant Ak$_2$ dimer.\cite{kristensen_quantum-state-sensitive_2022,kristensen_laser-induced_2023}

\subsection*{Data analysis}
The purpose of the data analysis is to extract the yields of the Ak$^+$He$_n$ ions as a function of the delay, $t$, between the pump and the probe pulse. Here $n$ denotes the number of He atoms attached to the \ce{Ak+} ion when it hits the detector. This is achieved by first identifying the peaks in the $m/q$-spectrum corresponding to the masses of \ce{Ak+He}$_n$ ions, i.e. $m/q = (m_{\rm{Ak}} + n\cdot4)\ u/e$. The ions in each peak are binned together and a VMI image is generated using their $(v_x, v_y)$ values. The VMI image contains the projected velocity components and the full three-dimensional velocity distribution is retrieved using the MEVIR algorithm.\cite{dick_inverting_2014} From this distribution, the yield of the Ak$^+$He$_n$ ions, Y$_n$, is obtained by summing all ions inside a restricted velocity range, see Methods section of Ref.\cite{albrechtsen_observing_2023} for details. This process is repeated for all delays measured between 0.2 ps and 20 ps, resulting in the main observable, $Y_n(t)$, called the time-dependent yield.

\subsection*{Computational details}
As in our previous investigation on the Na$^{+}$-doped helium clusters,\cite{albrechtsen_femtosecond-and-atom-resolved_2025} we have built the global PESs describing both the two-body (2B) interactions between the alkali ion Ak$^+$ and the He atoms and the corresponding three-body (3B) non-covalent contributions, whereas for describing the 2B He--He interaction we have employed the potential reported by Aziz.\cite{aziz_an_examination_1991} The 2B Li$^+$--He and Na$^+$--He interactions were already reported in Ref.\cite{rastogi_lithium_2018} and Ref. \cite{albrechtsen_femtosecond-and-atom-resolved_2025}, respectively. In the case of the 2B K$^+$--He interaction, reference {\it ab initio} energies have been here computed for a series of interparticle distances at the coupled cluster with single, double and perturbative triple excitations [CCSD(T)] level, obtained by using the Molpro2012.1 package.\cite{MOLPRO}

In particular, accurate counterpoise corrected K$^{+}$--He interaction energies have been computed by using the def2-AQZVPP \cite{Weigend} and d-aug-cc-pV6Z\cite{Dunning} basis sets for K$^{+}$ and He, respectively. We have checked that the computed interaction energies are well-converged by verifying that they differ by less than 1\% from those obtained in the minimum region with the def2-AQZVPP/d-aug-cc-pV5Z set.

The obtained ab initio results have been represented by the analytical improved Lennard-Jones (ILJ) formulation:\cite{pirani_beyond_2008}
\begin{equation}
\label{ILJ}
V(r) = \epsilon
\left[
{\frac{m}{n(r) \! - \!m}} \left( \frac{r_m}{r} \right)^{n(r)}
\! - {\frac{n(r)}{n(r) - m}} \left(\frac{r_m}{r} \right)^{m}
\right]
\end{equation}
\noindent
where $\epsilon$ is the potential depth, $r_m$ the position of the minimum and $n(r)$ is defined as:\cite{pirani_beyond_2008}
\begin{equation}
\label{nILJ}
n(r) = \beta + 4 \left( \frac{r}{r_m} \right)^2.
\end{equation}
A fine tuning of the parameters of the ILJ expression has been carried out by exploiting the comparison with the reference CCSD(T) interaction energies, as shown in Fig. S1 included in the Supplementary Information, where quite good agreement between the {\it ab initio} estimations and their analytical representation is seen. Such parameters are reported in Table \ref{tbl:PESparams} in comparison with the values corresponding to the Li$^+$--He\cite{rastogi_lithium_2018} and Na$^+$--He\cite{albrechtsen_femtosecond-and-atom-resolved_2025} cases. It is of relevance to note that for all three Ak$^+$--He systems, the adopted parameters are in very good agreement with those predicted by correlation formulas (see Cappelletti {\it et al.} \cite{capelletti_generalization_1991}), given in parenthesis in Table \ref{tbl:PESparams}, that exclusively exploit polarizability and charge of the interacting  partners. This finding confirms the pure non-covalent nature of the interaction, being only determined by the balance of exchange (size) repulsion with induction plus dispersion attraction. Moreover, the smaller stability of K$^+$--He is mostly determined by the largest exchange repulsion.

\begin{table}[h]
\small
  \caption{\ ILJ parameters for the 2B potential for the present case K$^+$--He, in comparison with those for Li$^+$--He and Na$^+$--He previously reported in Refs \cite{rastogi_lithium_2018} and Ref. \cite{albrechtsen_femtosecond-and-atom-resolved_2025}, respectively. The binding energy, $\epsilon$, is measured in meV; the equilibrium distance, $r_m$, in \AA. $\beta$ and $m$ are dimensionless. In parenthesis we have included predicted vaues from correlation formulas of Ref. \cite{capelletti_generalization_1991}}\label{tbl:PESparams}
  \begin{tabular*}{\columnwidth}{@{\extracolsep{\fill}}lllll}
    \hline
     & $\epsilon$ (meV) & $r_m$ (\AA) & $\beta$ & $m$ \\
    \hline
    Li$^+$--He \cite{rastogi_lithium_2018}
    & 81.3 (82) & 1.90 (1.91) & 4.2 & 4 \\
    Na$^+$--He \cite{albrechtsen_femtosecond-and-atom-resolved_2025}
    & 43.0 (45) & 2.31 (2.27) & 6.5 & 4  \\
    K$^+$--He & 22.5 (22) & 2.84 (2.83) & 8.5 & 4  \\
    \hline
    \end{tabular*}
\end{table}
\noindent Our investigation also comprises the description of the 3B non-covalent interaction to be added to the pairwise 2B contributions from the He--He and Ak$^+$--He terms. In this work, we have followed the same procedure as in previous studies of similar systems \cite{albrechtsen_femtosecond-and-atom-resolved_2025, rastogi_lithium_2018} given by:
\begin{eqnarray}
\label{eq:3B}
V_{\rm 3B} & = & -\frac{\alpha^2}{2}
\left[
{\frac{3 r_j}{2}} g_3(r_i) g_5(r_{ij}) + {\frac{3 r_i}{2}} g_3(r_j) g_5(r_{ij})
\right. \nonumber \\
& - & {\frac{1}{2}} g_3(r_i) g_3(r_j) g_1(r_{ij})
-3 g_1(r_i) g_1(r_j) g_5(r_{ij}) \nonumber \\
& - & \left.
g_1(r_i) g_3(r_j) g_3(r_{ij}) - g_3(r_i) g_1(r_j) g_3(r_{ij})
\right]
\end{eqnarray}
\noindent
which represents the main charge-induced dipole-induced dipole 3B term, and where $r_i$ corresponds to the distance of the $i-$th He atom and the Ak$^+$ ion; $r_{ij}$ describes the distance between the $i-$th and $j-$th He atoms and $\alpha =$ 1.45 , 1.42 or 1.31 $a_0^3$ for K$^+$, Na$^+$  and Li$^+$ respectively. For the $g_n(r_i)$ functions, we have employed here for K$^+$ the  same choice as for Na$^+$:\cite{albrechtsen_femtosecond-and-atom-resolved_2025} $g_n(r_i) = 1 / r_i^n$, while in the study on Li$^+$ they adopt the form $g_n(r_i) = f_n(r_i) / r_i^n$, where $f_n(r)$ are damping functions defined as:
\begin{equation}
\label{eq:damp}
f_n(r) = 1 - {\rm exp}(-br) \sum^n_{k = 0} \frac{[br]^k}{k!}
\end{equation}
\noindent
with the parameter $b$ equals to 2.9 $a_0^{-1}$ or 3.2 $a_0^{-1}$ for the Li$^+$--He and He--He interactions, respectively.

With the above described PESs, we have employed a PIMC method \cite{rodriguez-cantano_path-integral_2016} to calculate: (i) evaporation energies, defined as $E_{\rm evap}(N) = E_{\rm bind}(N) - E_{\rm bind}(N-1)$, where $E_{\rm bind}(N)$ stands for the binding energy of the Ak$^+$He$_N$ cluster, and (ii) the structures for clusters of different sizes. The PIMC method has already been described many times before in previous applications,\cite{rastogi_lithium_2018,albrechtsen_femtosecond-and-atom-resolved_2025,rodriguez-cantano_path-integral_2016,rodriguez-cantano_a-configurational_2015} so here, we just mention that a thermodynamic estimator \cite{rodriguez-cantano_path-integral_2016} has been used and a variable number of quantum beads between 600 and 1200 have been considered in the calculations. The PIMC simulations are initiated from either classically optimized or selected configurations.

\section*{Results and discussion}
First, we present the theoretical results, which both motivate the subsequent analysis of the experimental Ak$^+$He$_n$ yields and help interpret the observed differences between the three alkali ion species.
\subsection*{Evaporation energies and characteristics of Ak$^+$He$_N$ complexes}\label{sec_evaporationEnergies}
\begin{figure}[h]
  \centering
    \includegraphics[width=\columnwidth]{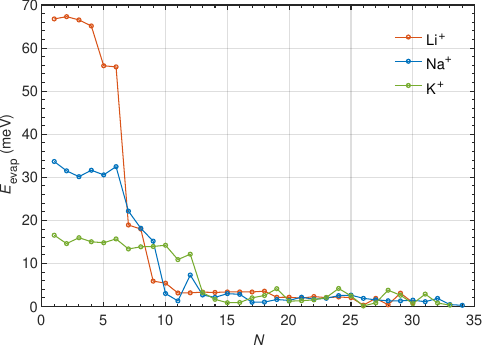}
    \caption{The evaporation energy, $E_{\text{evap}}(N)$, of the Ak$^+$He$_N$ complexes calculated by PIMC. Results for Li$^+$ (red), Na$^+$ (blue) and K$^+$ (green). Values for Li$^+$ are from Ref.\cite{rastogi_lithium_2018}, values for Na$^+$ are from Ref.\cite{albrechtsen_femtosecond-and-atom-resolved_2025} and K$^+$ results are from this paper.}\label{fgr:evaporation_energies}
\end{figure}

Figure \ref{fgr:evaporation_energies} shows the calculated evaporation energies, $E_{\text{evap}}$, of Li$^+$He$_N$, Na$^+$He$_N$ and K$^+$He$_N$ for the first 30, 34 and 33 He atoms, respectively. Here, we use $N$ for the number of He atoms bound to the \ce{Ak+} ion to make a distinction from the Ak$^+$He$_n$ ions recorded by the detector in the experiment.
The evaporation energies for all three alkali ions are largest for the first few helium atoms and then decrease at larger values of $N$. Li$^+$ has the highest evaporation energies for the first helium atoms followed by Na$^+$ and then K$^+$, i.e. a trend that reflects the binding energies for the corresponding \ce{Ak+}--He pair potentials, see Table \ref{tbl:PESparams}. Significant features are found at $N=6$ for Li$^+$ and Na$^+$. Beyond this size, a steep decrease in $E_{\text{evap}}$ is observed for Li$^+$ while Na$^+$ exhibits a gradual reduction reaching a plateau at $N$ = 10. For K$^+$, the first 12 He attachments have a similar $E_{\text{evap}}$ before decreasing.

For Na$^+$, the evaporation energy exhibits a local maximum at $N=12$. This particular size has been found to correspond to a stable configuration in other systems as well such as Ca$^{2+}$He$_{12}$,\cite{zunzunegui-bru_observation_2023} Ar$^+$He$_{12}$,\cite{tramonto_path_integral_2015,bartl_on_the_size_2014} Ho$^{2+}$He$_{12}$\cite{foitzik_formation_2025} or Ag$^+$He$_{12}$.\cite{doppner_ion_2007} Previous theoretical investigations on Na$^+$He$_N$ complexes concluded that the closure of the first solvation shell of He atoms around the ion actually occurs for $N=12$.\cite{coccia_bosonic_2007, galli_path_integral_2011, issaoui_theoretical_2014, laajimi_structure_2020, rossi_alkali_2004, paolini_ground_state_2007} The only exception of this apparent overall agreement comes from the estimate by integration of the radial density profile performed by Galli {\it et al.}\cite{galli_pure_2001} and the abrupt drop of the experimental ion yield of the work by An der Lan {\it et al.},\cite{an_der_lan_solvation_2012} which may suggest the relevance of $N=9$. In this sense, one might argue that the decrease seen at $N=9$ of $E_{\text{evap}}$ shown in Fig. \ref{fgr:evaporation_energies} could be understood as a special feature for such a configuration. In the case of K$^+$, there is general agreement that there are 14 He atoms in the first solvation shell,\cite{coccia_bosonic_2007, galli_pure_2001, an_der_lan_solvation_2012} although Galli {\it et al.} \cite{galli_path_integral_2011} and Rossi {\it et al.} \cite{rossi_alkali_2004} suggest that the number could be 15. Recent density functional theory calculations in helium nanodroplets by Garc\'{\i}a {\it et al.}\cite{garcia-alfonso_time-resolved_2025} systematically predict a larger number of He atoms for the first solvation shell for all the alkali ions than those reported in the rest of the literature.

\begin{figure}[h]
  \centering
    \includegraphics[width=\columnwidth]{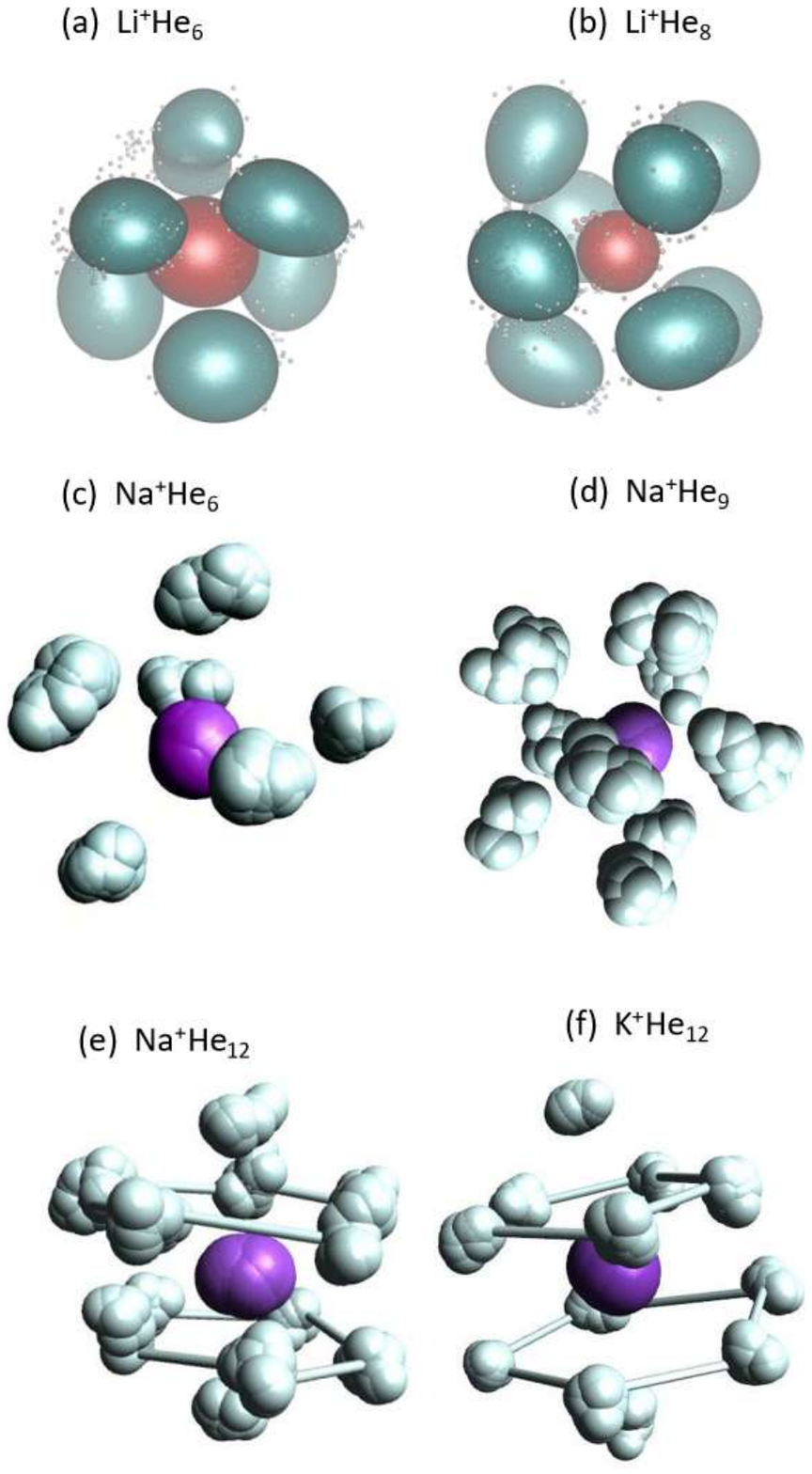}
    \caption{Structures of relevant solvation complexes with evaporation energies shown in Figure \ref{fgr:evaporation_energies} and obtained by considering either an average (Li$^+$) or a limited value (Na$^+$ and K$^+$) of quantum beads for a better ilustration. (a) and (b) are PIMC results adapted from Ref. \cite{rastogi_lithium_2018} for Li$^+$He$_6$ and Li$^+$He$_8$, respectively; (c), (d) and (e) have been estimated from the calculation reported Ref. \cite{albrechtsen_femtosecond-and-atom-resolved_2025} for Na$^+$He$_6$ and Na$^+$He$_{12}$, respectively and finally (f) is the structure from the present PIMC calculation for K$^+$He$_{12}$.}
    \label{fgr:structures}
\end{figure}

For all Ak$^+$He$_N$ complexes with $N > 12$, subsequent helium attachments exhibit roughly a constant evaporation energy ($< 5 $ meV) independent of the alkali species. This is ascribed to the shielding effect of the first solvation shell of helium atoms around the ion. This shielding causes the next helium atoms to experience a similar ionic attraction irrespective of the ion species. This makes the first helium attachments particularly important for observing differences between the solvation of the alkali species.

To further corroborate this analysis, we consider the structures of the Ak$^+$He$_N$ complexes for the values of $N$ where a significant change in the evaporation energy is observed.
These are displayed in Fig. \ref{fgr:structures}. The structures show an important interplay between charge density and ion size. Li$^+$ is the smallest ion with the highest charge density and will therefore bind the strongest to helium, but due to the small size, only a few helium atoms can bind independently before He--He repulsion becomes noticeable. For K$^+$, a lot of helium atoms can attach before any strong repulsion between the He atoms is observed. However, due to the smaller charge density, these helium atoms are not very strongly bound. Na$^+$ places somewhere between these two systems. From a static point of view, we can then already see significant differences in the helium solvation properties of the three alkali ions, which then become less pronounced as the clusters grow in size. From Fig. \ref{fgr:structures}, it is clear that only the structures for Na$^+$He$_{12}$ and K$^+$He$_{12}$, in the bottom panels (e) and (f), exhibit a closed, well-ordered icosahedral geometry.

\subsection*{Experimental time-dependent ion yields}
\begin{figure*}[h]
\centering
  \includegraphics[width=\textwidth]{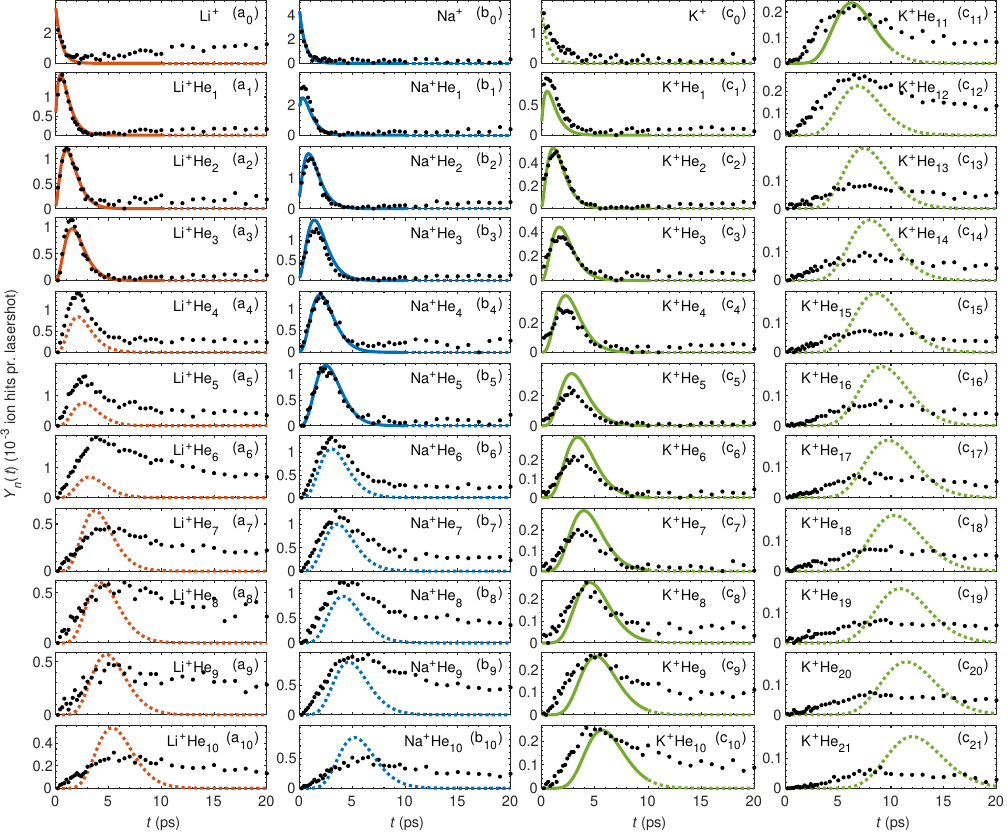}
  \caption{Time-dependent ion yields, $Y_n(t)$. Left column of panels (red): Li$^+$He$_n$. Second left column of panels (blue): Na$^+$He$_n$. Two rightmost columns of panels (green): K$^+$He$_n$. Black dots are experimental data, full lines are the results from a fit to Eq. \ref{eq:poiss} and dashed lines are extrapolated values of the fit beyond the fitted range.}\label{fgr:yields}
\end{figure*}

Figure \ref{fgr:yields} displays the Ak$^+$He$_n$ yields as a function of time for the three alkali species. For \ce{Li+} (first column) and \ce{Na+} (second column), $Y_n(t)$ are shown for $0 \leq n \leq 10$ but data were recorded up to $n_\text{max} = 29$ (\ce{Li+}) and $n_\text{max} = 25$ (\ce{Na+}), the maximum value of $n$ being limited by overlap with Xe$^+$He$_k$ ions in the $m/q$-spectrum. In the case of \ce{K+}, all $Y_n(t)$ up to $n_\text{max} = 21$ are displayed (third and fourth columns). The $Y_n(t)$  traces show a similar trend for the three alkali species. Notably, $Y_0(t)$, i.e. the yield of the bare alkali ions [panels (a$_0$), (b$_0$) and (c$_0$)] gradually decreases whereas for $n \geq 1$, the ion yields reach a maximum followed by a gradual decrease. The progressive shifting of the maximum to larger times for increasing $n$ illustrates directly that He atoms gradually bind to the alkali ions.

Motivated by theoretical simulations\cite{garcia-alfonso_time-resolved_2024} and our previous work,\cite{albrechtsen_femtosecond-and-atom-resolved_2025} we employ a Poissonian model to fit $Y_n(t)$ for the three alkali ions:
\begin{equation}
\label{eq:poiss}
  f(t;n) = A\frac{(r(t+t_{\text{lift}}))^n\exp^{-r(t+t_{\text{lift}})}}{n!}
\end{equation}
where $A$ is an amplitude, $r$ is the binding rate, $t_{\text{lift}}$ is a correction factor of the experimental time delay due to effects in the probe process,\cite{albrechtsen_femtosecond-and-atom-resolved_2025} and $t$ is the pump-probe delay. We apply a global fit of Eq. (\ref{eq:poiss}) to $Y_n(t)$, but exclude some values of $(t;n)$ in the fit. This is because the Poissonian model only applies when the binding rate is constant and the binding of helium atoms occurs independently. When $n$ grows larger, the He--He repulsion plays an increasing role as seen in Fig. \ref{fgr:evaporation_energies} indicating that helium atoms no longer bind independently. Therefore, we only apply the fit to the Ak$^+$He$_n$ ions with the lowest $n$-values where the $E_{\text{evap}}$ is approximately constant. Thus, based on the results shown in Fig. \ref{fgr:evaporation_energies}, we only include $n$ = 0-3 for Li$^+$, $n$ = 0-5 for Na$^+$ and $n$ = 1-11 for K$^+$. The K$^+$ signal is excluded because it contains contaminants from ionization of the K$_2$ dimer. Furthermore, only the first 10 ps of the experimental data are included because the later dynamics are significantly influenced by the dissociation of large Ak$^+$He$_n$ complexes.\cite{albrechtsen_femtosecond-and-atom-resolved_2025}.

The result of the fit of Eq. (\ref{eq:poiss}) to each alkali ion measurement is shown by the full lines in Fig. \ref{fgr:yields} and the fit parameters are given in Table \ref{solvation_fitting_results}. The dashed lines in Fig. \ref{fgr:yields} illustrate the result of the fit beyond the range included in the fit ($t > 10$~ps and the alkali-specific $n$-ranges).
\begin{table}[h]
\small
  \caption{\ Parameters resulting from fitting Eq. (\ref{eq:poiss}) to the experimental data shown in Fig. \ref{fgr:yields}, for the three different alkali ions, including the range of fitted complexes sizes ($n_{\text{inc}}$), the fitted helium binding rate
  ($r$) and the lift-off correction factor (t$_{\text{lift}}$)}\label{solvation_fitting_results}
  \begin{tabular*}{\columnwidth}{@{\extracolsep{\fill}}lllll}
    \hline
    Alkali & $n_{\text{inc}}$ (min-max) & $r$ (He ps$^{-1}$) & $t_{\text{lift}}$ (ps)\\
    \hline
    Li & 0-3 & 1.8 $\pm$ 0.2 & 0.09 $\pm$ 0.09\\
    Na & 0-5 & 1.8 $\pm$ 0.1 & 0.24 $\pm$ 0.06\\
    K & 1-11 & 1.7 $\pm$ 0.1 & 0.08 $\pm$ 0.08\\
    \hline
  \end{tabular*}
\end{table}
First, we note that the Poissonian model agrees well with the experimental results in the respective fitting ranges for the three ions. Going beyond these ranges, the  $Y_n(t)$ curves deviate significantly from the corresponding Poisson curves, as seen for the case of \ce{Li+}, Fig. \ref{fgr:yields}(a$_4$)-(a$_{10}$), \ce{Na+}, Fig. \ref{fgr:yields}(b$_6$)-(b$_{10}$), and \ce{K+}, Fig. \ref{fgr:yields}(c$_{12}$)-(c$_{21}$). The reasons for the deviations were already discussed for \ce{Na+} in the two previous works, Refs.\cite{albrechtsen_observing_2023,albrechtsen_femtosecond-and-atom-resolved_2025}. Here, we briefly account for the \ce{Li+} and \ce{K+} results.

In the \ce{Li+} case, Fig. \ref{fgr:yields}(a$_4$) shows that $Y_4(t)$ is significantly higher than the Poisson curve. In line with the discussions in Refs. \cite{albrechtsen_observing_2023,albrechtsen_femtosecond-and-atom-resolved_2025}, we ascribe this to dissociation of \ce{Li^+He_$n$} with $n \geq 5$ because of the drop in $E_{\text{evap}}$ from $n = 4$ to 5, see Fig.\ref{fgr:evaporation_energies}. This effect is even more pronounced for $Y_6(t)$ where the maximum is almost a factor of 4 higher than the maximum in the Poisson curve. The explanation is found in the strong drop of $E_{\text{evap}}$ (55 to 18 meV) from $N = 6$ to 7, Fig. \ref{fgr:evaporation_energies}, that causes \ce{Li^+He_7} or larger complexes leaving from the droplet to likely loose a He atom (or more He atoms) and thus be detected as a \ce{Li^+He_6} ion. Note that the position of the maximum of $Y_6(t)$ appears later than the maximum of the corresponding Poisson curve corroborating that the \ce{Li^+He_6} ions originate from larger complexes that were formed later but lost He atoms after departing from the droplet surface. The strong $Y_6(t)$ signal indicates that Li$^+$He$_6$ most likely is a filled solvation shell or, at least, a significantly stable structure compared to larger complexes as predicted by theory.

In the \ce{K+} case, the $Y_n(t)$ and Poisson curves agree quite well in the fitting range, i.e. up to $n = 11$. This reflects, we believe, that $E_{\text{evap}}$ is essentially constant in the range $N = 1-12$. The slightly less good agreement between the $Y_n(t)$ and Poisson curves compared to the \ce{Li+} an \ce{Na+} cases is likely a result of the significantly lower $E_{\text{evap}}$, which could causes some dissociation to occur in the fitted $n$-range. For $Y_{12}(t)$, the maximum rises above that of the Poisson curve, whereas for $Y_{13}(t)$ the maximum falls much below that of the Poisson curve. This behavior, caused by the drop in $E_{\text{evap}}$  from $\sim$12 meV at $N= 12$ to 3 meV  at $N = 13$, is similar to that seen in the \ce{Li+} data from $Y_{6}(t)$ to $Y_{7}(t)$  and in the \ce{Na+} data from $Y_{9}(t)$ to $Y_{10}(t)$. As for the \ce{Li+} and \ce{Na+} cases, the drop in $E_{\text{evap}}$ means that one or more He atoms evaporate from K$^+$He$_n$ with $n > 12$, leading to a depletion of $Y_n(t)$ and a corresponding pile-up in $Y_{12}(t)$.

Table \ref{solvation_fitting_results} shows that the He binding rate, $r$, is about the same for the three alkali species with a value of $\sim$1.7-1.8 $\pm$ 0.1 He per ps. Regarding comparison to theoretical calculations, we refer to the very recent work of García-Alfonso et al. where time-dependent density-functional theory was used to simulate the alkali ion solvation in nanodroplets consisting of 2000 He atoms.\cite{garcia-alfonso_time-resolved_2025} The simulations were carried out for \ce{Na+}, \ce{K+}, \ce{Rb+} and \ce{Cs+} and thanks to a smaller space step in the simulation grid, the results are expected to be more accurate than previous findings.\cite{garcia-alfonso_time-resolved_2024} In the case of \ce{Na+} and \ce{K+}, the theoretical results confirmed that the binding of the first five He atoms to the alkali ions occur at a constant rate. Furthermore, $r$ was found to be 1.3 and 1.0 He per ps for \ce{Na+} and \ce{K+}, respectively. In our previous work on \ce{Na+}, $r$-values of 2.04$~\pm$~0.13, 1.84~$\pm$~0.09 and 1.65~$\pm$~0.09 were determined for droplets with an average number of 9000, 5200, and 3600 He atoms, respectively. This decreasing trend of $r$ with droplet size indicates a good agreement between the experimental values and the simulated result obtained for a 2000 He atom droplets. For \ce{K+}, we also measured the binding rate for other droplet sizes namely droplets with an average number of 9000, 5200 and 3600 He atoms. Here, we found a similar trend with an obtained binding rate of 1.9$~\pm$~0.1, 1.7$~\pm$~0.1 and 1.6$~\pm$~0.1 He per ps, again pointing to a good agreement between the experimental and simulated binding rates.

\subsection*{Number distribution of ions}
\begin{figure*}[h]
\centering
  \includegraphics[width=\textwidth]{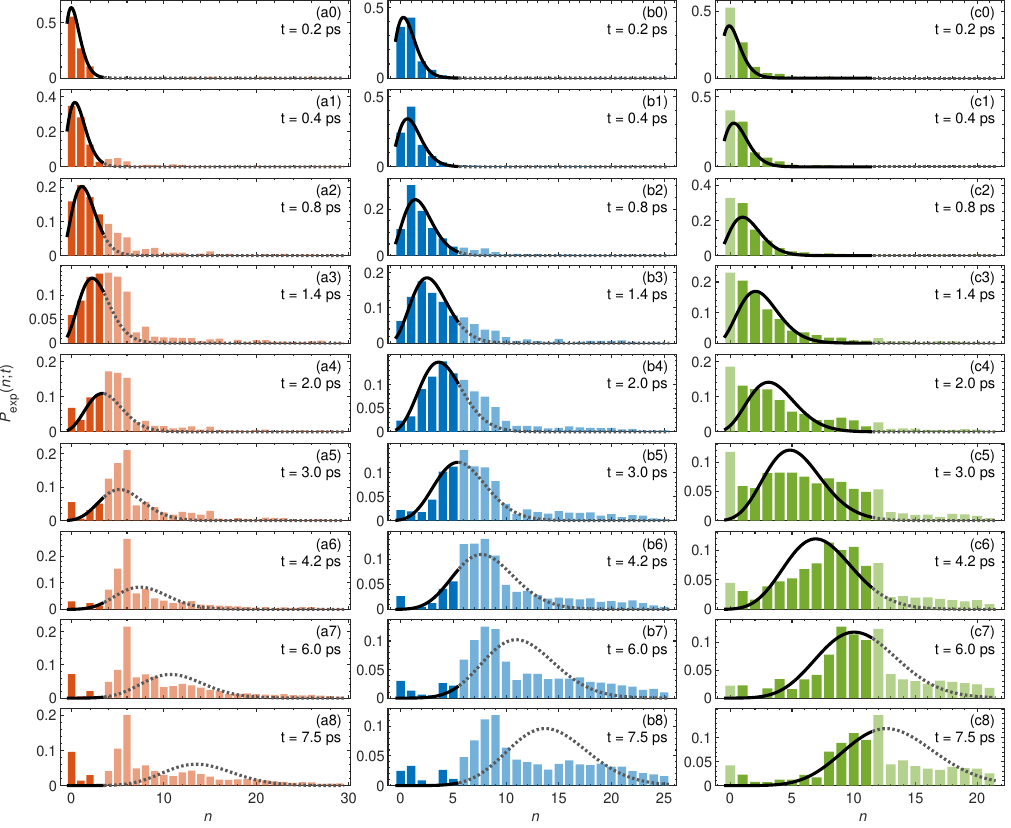}
  \caption{Number distributions, $P_{\text{exp}}(n;t)$, Left panels (red): Li$^+$He$_n$. Central panels (blue): Na$^+$He$_n$. Right panels (green): K$^+$He$_n$. Filled bars: experimental data included in the fit, greyed-out bars: data excluded in the fit, filled black lines: result of the fit, dashed lines: result of the fit extrapolated beyond the fitted range.}
  \label{fgr:distributions}
\end{figure*}

From the time-dependent ion yields, $Y_n(t)$, we conversely determine a distribution of ion yields, $Y_t(n)$ at each time delay, $t$. From this, we calculate the normalized number distribution, $P_{\text{exp}}(n;t)$, as:
\begin{equation}
\label{number-distribution}
P_{\text{exp}}(n;t) =  \frac{Y_t(n)}{\sum^{n_{max}}_{i = 0}Y_t(i)}.
\end{equation}
\noindent
Figure \ref{fgr:distributions} displays $P_{\text{exp}}(n;t)$ at nine selected times for \ce{Li+}, \ce{Na+} and \ce{K+}. In common for the three ion species, it is seen that at the earliest times, \ce{Ak+} and \ce{Ak+He} dominates $P_{\text{exp}}(n;t)$. This observation is expected since the alkali ion has not had much time to interact with and effectively bind He atoms. At longer times, the number distributions broaden and their weights displace to larger $n$ values. This behavior illustrates that the solvation process of the \ce{Ak+} ions, i.e. the gradual binding of He atoms, is stochastic.

Figure \ref{fgr:distributions} also depicts the outcome of the fit applied to $P_{\text{exp}}(n;t)$, i.e. Eq. \ref{eq:poiss} using the parameters listed in Table \ref{solvation_fitting_results}. The fitted results (full black curves) agree well with the experimental data in the fitting ranges and, similar to the $Y_n(t)$ curves in Fig. \ref{fgr:yields}, deviations appear beyond these ranges, where the fit is extrapolated (black dotted curves). In the \ce{Li+} case, we observe that the amplitudes of $P_{\text{exp}}(4)$, $P_{\text{exp}}(5)$ and $P_{\text{exp}}(6)$ at early times, $t$ = 0.8 ps - 3.0 ps, are significantly larger than the Poisson curves, see Fig. \ref{fgr:distributions}(a$_2$)-(a$_5$). Equivalently, the $Y_4(t)$, $Y_5(t)$ and $Y_6(t)$ curves lie significantly above the Poisson curves, see Fig. \ref{fgr:yields}(a$_4$)-(a$_6$). These observations could be due to many-body binding processes where several He atoms attach to the ion simultaneously. Such non-Poissonian processes may be particularly important for the solvation of the \ce{Li+} ion because the interaction with the He atoms is the largest of the three alkali ions. Finally, for all three ion species, we see a clear edge in the number distributions, i.e. a $n$-value after which the amplitude significantly decreases. For \ce{Li+} the $n$-value is 6 observed when $t \geq 1.4 $~ps, for \ce{Na+} it is 9 observed when $t \geq 2.0 $~ps and for \ce{K+} it is 12 observed when $t \geq 3.0 $~ps. These numbers reflect the $N$-values where $E_{\text{evap}}(N)$ makes a steep drop, Fig. \ref{fgr:evaporation_energies}, as mentioned in the discussion of the $Y_n(t)$ curves. 

\subsection*{Energy dissipation}
When the \ce{Ak+} ions solvate in the helium droplets, energy is removed from the local region around \ce{Ak+}. Once the solvation is complete, an energy of 604~meV for \ce{Li+}, 385~meV for \ce{Na+} and 287~meV for \ce{K+} has been dissipated.\cite{garcia-alfonso_time-resolved_2024} This energy, termed the sinking energy, $E_\text{sink}$ in Ref.\cite{garcia-alfonso_time-resolved_2024}, is the energy difference between \ce{Ak+} at its initial position at the surface and when it is fully solvated in the droplet interior. In our two previous works,\cite{albrechtsen_observing_2023,albrechtsen_femtosecond-and-atom-resolved_2025} we showed that the mean energy dissipated at time $t$, $\langle E_{\text{disp}} \rangle(t)$, can be determined from the distribution of the Ak$^+$He$_n$ ions recorded at $t$, Fig. \ref{fgr:distributions}, and the binding energy of each Ak$^+$He$_n$ ion, $E_{\text{bind}}(n)$:
\begin{equation}\label{eq:dissipation}
    \begin{aligned}
        &\sum_{i=1}^{n_{\text{max}}}P_{\text{exp}}(i;t)(-E_{\text{bind}}(i-1)+
        E_{\text{init}}(\text{Ak}^{+}))\leq\langle E_{\text{disp}} \rangle(t)\\
        &<\sum_{i=1}^{n_{\text{max}}}P_{\text{exp}}(i;t)(-E_{\text{bind}}(i)+E_{\text{init}}(\text{Ak}^{+})).
    \end{aligned}
\end{equation}
Here $n_{\text{max}}$ is the largest $n$-value measured, as mentioned above, $E_{\text{bind}}(i)$ is obtained as the sum of the negative evaporation energies, Fig. \ref{fgr:evaporation_energies}, from 1 to $i$ and $E_{\text{init}}$(\ce{Ak+}) is the initial energy of the Ak$^+$ ion at the surface of a 2000 He droplet.\cite{garcia-alfonso_time-resolved_2024} The reason behind Eq. \ref{eq:dissipation} is that when an Ak$^+$He$_N$ ion leaves the droplet, it quickly gets rid of any internal energy by shedding the number of He atoms energetically available and, thus, bring the internal energy below the dissociation energy of the now reduced-size ion complex Ak$^+$He$_n$. This ion flies intact to the detector and from its size, i.e. $n$, which we record, the energy dissipated at the droplet, $E_{\text{disp}}$, is determined as $-E_{\text{bind}}(n)$. At any time, a distribution of Ak$^+$He$_n$ ions is recorded, Fig. \ref{fgr:distributions}, so $E_{\text{disp}}$ must be weighted by $P_{\text{exp}}(n;t)$ as expressed in Eq. \ref{eq:dissipation}. The interval of $\langle E_{\text{disp}} \rangle(t)$ given in Eq. \ref{eq:dissipation} reflects that an \ce{Ak^+He_n} ion detected may retain internal energy up to its evaporation energy.

\begin{figure}[h]
\centering
  \includegraphics[width=\columnwidth]{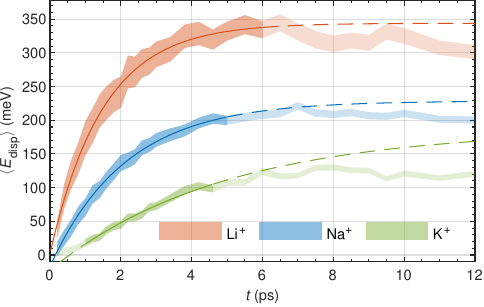}
  \caption{Mean dissipated energy as a function of time for Li$^+$ (red), Na$^+$ (blue) and K$^+$ (green). The dark-colored regions indicate values included in the fit of Eq. (\ref{eqn:newt}). The fit is shown as the full colored lines, while the dashed colored lines represent the fit extrapolated beyond the fitted time ranges}\label{fgr:disspation_energy}
\end{figure}

The outcome of applying Eq. \ref{eq:dissipation} to the experimental data, displayed in Fig. \ref{fgr:disspation_energy}, shows a similar trend of $\langle E_{\text{disp}} \rangle(t)$ for the three ion species. Initially, $\langle E_{\text{disp}} \rangle(t)$ rises steeply after which it flatten off and asymptotically reaches a maximum value. Similar to previous work,\cite{albrechtsen_femtosecond-and-atom-resolved_2025} we fit the following function to the experimental results:
\begin{equation}
    E_{\text{disp}}^{\text{Newt}}(t) =
    E_{\text{disp}}(\infty)\left(1-\exp\left(-\frac{(t+\Delta t_{\text{disp}})}{\tau_{\text{disp}}}\right)\right)
    \label{eqn:newt}
\end{equation}
where $E_{\text{disp}}(\infty)$ is the asymptotic value of the dissipation energy, $\tau_{\text{disp}}$ is the time constant and $\Delta t_{\text{disp}}$ is a time offset. Equation (\ref{eqn:newt}) is motivated by Newton's law of cooling, which states that the rate of heat transfer between a hot (or energetic) body and the surrounding environment is proportional to the internal energy of the body. We only apply the fit to the time range where we are certain that all Ak$^+$He$_n$ ions arriving at the detector can be identified and recorded, i.e. when $n \leq n_{max}$. Based on the number distributions in Fig. \ref{fgr:distributions}, we consider this fulfilled in the ranges 0-6.0 ps for \ce{Li+}, 0-5.0 ps for \ce{Na+} and 0-4.5 ps for \ce{K+}. At longer times, there will be Ak$^+$He$_n$ ions with $n > n_{max}$ and those we are not able to detect due to the overlap with Xe$^+$He$_k$ ions. Therefore, we expect that $\langle E_{\text{disp}} \rangle$ underestimates the real energy dissipated, an effect that increases with time since the progression of the solvation process leads to gradually larger Ak$^+$He$_n$ ions arriving at the detector.

Figure \ref{fgr:disspation_energy} shows that the fits of Eq. (\ref{eqn:newt}) agree well with the experimental data in the time ranges selected. For \ce{K+}, there is a minor deviation during the first ps, which we believe is caused by a small contribution to the \ce{K+} and \ce{K^+He} signal from dissociative ionization of \ce{K2}.
\begin{table}[h]
\small
  \caption{\ Parameters obtained by fitting Eq .(\ref{eqn:newt}) to the experimental data in Fig. \ref{fgr:disspation_energy} for all three alkali ions, including the energy dissipation time constant ($\tau_{\text{disp}}$), energy dissipation correction-factor ($\Delta t_{\text{disp}}$) and asymptotic dissipated energy value ($E_{\text{disp}}(\infty)$)}\label{dissipation_results}
  \begin{tabular*}{0.48\textwidth}{@{\extracolsep{\fill}}llll}
    \hline
    Alkali & $\tau_{\text{disp}}$ (ps) & $\Delta t_{\text{disp}}$ (ps) & $E_{\text{disp}}(\infty)$ (meV) \\
    \hline
    Li & 1.5 $\pm$ 0.1 & 0.01 $\pm$ 0.05 & 344 $\pm$ 7 \\
    Na & 2.1 $\pm$ 0.2 & -0.15 $\pm$ 0.04 & 229 $\pm$ 8 \\
    K & 4.9 $\pm$ 3.7 & -0.55 $\pm$ 0.25 & 187 $\pm$ 92  \\
    \hline
  \end{tabular*}
\end{table}
The parameters of the fits are given in Table \ref{dissipation_results}. Firstly, the decreasing value of $\tau_{\text{disp}}$ as the alkali ions become larger reflects the obvious fact from the experimental curves in Fig. \ref{fgr:disspation_energy}, that energy dissipates fastest in the Li$^+$ case followed by Na$^+$ and then K$^+$. This observation makes sense because we found that the helium binding rates were similar for the three alkali ions and since the order of the total energy released in the solvation process, i.e. $E_\text{sink}$ is $E_\text{sink}$(\ce{Li+}) $>$ $E_\text{sink}$(\ce{Na+}) $>$ $E_\text{sink}$(\ce{K+}), then the energy dissipation rate must be in the same order. Secondly, the values of $E_{\text{disp}}(\infty)$ are $\sim$57~\%, $\sim$59~\% and $\sim$65~\% of the values of $E_\text{sink}$ given above. This experimental underestimate reflects the limitation of the experiment in terms of not being able to measure Ak$^+$He$_n$ ions with $n > n_{max}$ when a Xe atom is used as the interior dopant to create the repeller ion. It may be possible to lift this limitation and extend the range of Ak$^+$He$_n$ ions detected to larger $n$-values by choosing another interior dopant. For instance, \ce{SF6} is a possible candidate.

\section*{Conclusion}
We investigated the solvation dynamics of single \ce{Li+}, \ce{Na+} and \ce{K+} ions in helium nanodroplets, on the natural femtosecond and picosecond timescale and with atomic number-resolution. A fs pump pulse initiated solvation by ionizing the alkali atom, residing at the droplet surface and the ensuing dynamics was inferred by ionizing a Xe atom in the center of the droplet, by a delayed fs probe pulse, and recording the yield of Ak$^+$He$_n$ ions, repelled from the droplet, as a function of time. We found that the binding of individual He atoms to the alkali ions is well-described by a Poisson process for the first 3, 5 and 11 He atoms for \ce{Li+}, \ce{Na+} and \ce{K+}, respectively. From the Poisson analysis of the time-dependent Ak$^+$He$_n$ ion yields, we determined a binding rate of 1.8~$\pm$~0.1, 1.8~$\pm$~0.1 and 1.7~$\pm$~0.1 He per ps, respectively for the three ions. For \ce{Na+} and \ce{K+}, these values could be compared to very recent results from time-dependent density functional simulations, revealing good agreement when accounting for differences in the droplet sizes explored experimentally and theoretically. Furthermore, in the number distributions of the Ak$^+$He$_n$ ions, we observe pronounced edge effects at $n$ = 6 for \ce{Li+}, $n$ = 9 for \ce{Na+} and $n$ = 12 for \ce{K+}. These numbers correspond to the values where the evaporation energy of the Ak$^+$He$_n$ ion makes a steep drop as predicted by our Path Integral Monte Carlo calculations.

We also determined the average energy dissipated from the local region of the \ce{Ak+} ions as a function of time. The energy dissipation follows Newton's law of cooling for the first $\sim$ 5 ps. The cooling was observed to occur fastest (slowest) for \ce{Li+} (\ce{K+}), which is also the ion that needs to dissipate the largest (smallest) amount of energy in the solvation process. Our method paves the way for similar studies of the singly as well as doubly charged alkaline-earth cations, which are still unexplored.

\section*{Author contributions}
Jeppe K. Christensen: conceptualization, formal analysis, investigation, methodology, validation, writing -- original draft, writing -- review \& editing. Simon H. Albrechtsen: conceptualization, formal analysis, investigation, methodology, visualization, writing -- review \& editing. Christian E. Petersen: formal analysis, investigation, methodology, writing -- review \& editing. Constant A. Schouder: conceptualization, investigation, methodology, writing -- review \& editing. Iker Sánchez-Pérez: investigation, methodology. Pedro Javier Carchi-Villalta: investigation, methodology. Massimiliano Bartolomei: investigation, methodology, writing -- review \& editing. Fernando Pirani: investigation, methodology. Tomás González-Lezana: funding acquisition, investigation, methodology, project administration, supervision, writing -- original draft, writing -- review \& editing. Henrik Stapelfeldt: conceptualization, formal analysis, funding acquisition, investigation, methodology, supervision, writing -- original draft, writing -- review \& editing. 

\section*{Conflicts of interest}
There are no conflicts to declare.

\section*{Data availability}
Data for this article, including calculated evaporation energies shown in Fig. 1, $Y_{n}(t)$ traces shown in Fig. 3, $P_{\text{exp}}(n;t)$ distributions shown in Fig. 4 and mean dissipated energies shown in Fig. 5, are available in the Supplementary Information.

\section*{Acknowledgements}
We thank Jan Thøgersen for his expert help in keeping the laser system in optimal condition. H.S. acknowledges the support from Villum Fonden through a Villum Investigator Grant No. 25886. This work has been supported by Grant No. MICIU/AIE/10.13039/501100011033 PID2024-155666NB-I00.



\balance


\bibliography{Manuscript-Christensen} 
\bibliographystyle{rsc} 
\end{document}